\documentclass[letterpaper, 10 pt, conference]{ieeeconf}  

\IEEEoverridecommandlockouts                              

\usepackage{amsthm}
\usepackage{graphicx}
\usepackage{amsmath}
\usepackage{amssymb}
\usepackage{xcolor}
\usepackage{cuted}
\usepackage{subfigure}
\usepackage{url}
\usepackage{algorithm}
\usepackage{algpseudocode}

\usepackage{siunitx}

 \theoremstyle{definition}

\newtheorem{assumption}{Assumption}

\newtheorem{safety}{Safety Condition}

\title{\LARGE \bf Deep Continuum Deformation Coordination and Optimization with Safety Guarantees}
\author{Harshvardhan Uppaluru and Hossein Rastgoftar%
\thanks{*This work has been supported by the National Science Foundation under Award Nos. 2133690 and 1914581.  }
\thanks{Authors are with Scalable Move And Resilient Traversability Lab, Aerospace and Mechanical Engineering Department, University of Arizona, Tucson - 85721, USA. {\tt\small Email: \{huppaluru, hrastgoftar\}@arizona.edu}}%
}

\begin{document}

\maketitle
\thispagestyle{empty}
\pagestyle{empty}

\begin{abstract}
In this paper, we develop and present a novel strategy for safe coordination of a large-scale multi-agent team with ``\textit{local deformation}" capabilities. Multi-agent coordination is defined by our proposed method as a multi-layer deformation problem specified as a Deep Neural Network (DNN) optimization problem. The proposed DNN consists of $p$ hidden layers, each of which contains artificial neurons representing unique agents. Furthermore, based on the desired positions of the agents of hidden layer $k$ ($k=1,\cdots,p-1$), the desired deformation of the agents of hidden layer $k + 1$ is planned. In contrast to the available neural network learning problems, our proposed neural network optimization receives time-invariant reference positions of the boundary agents as inputs and trains the weights based on the desired trajectory of the agent team configuration, where the weights are constrained by certain lower and upper bounds to ensure inter-agent collision avoidance. We simulate and provide the results of a large-scale quadcopter team coordination tracking a desired elliptical trajectory to validate the proposed approach.
\end{abstract}


\section{INTRODUCTION}
\label{sec:introduction}
First inspired by natural phenomena, formation flight and cooperative control in Multi-Agent Systems (MAS) have been fascinating and important areas of study for the past 20 years. Research into MAS has led to interesting theoretical problems and potential practical uses in a wide range of situations. MAS formation flight and cooperative control are achieved either in a centralized or decentralized manner. The centralized technique makes use of a central computer that controls every agent in the MAS. However, the decentralized technique, also known as distributed control, allows for computation onboard each agent, and information is shared across neighboring agents \cite{cao2012overview}. For cooperative multi-agent control, the decentralized method has many benefits, such as low operational costs, fewer system requirements, great robustness, strong adaptability, and flexible scalability.

\subsection{Related Work}
With applications ranging from surveillance \cite{du2017pursuing, da2017multi} to formation flying \cite{da2017multi, ha2017multi}, rescue missions \cite{altair2017decision}, wildlife monitoring and exploration \cite{witczuk2018exploring}, precision agriculture \cite{tsouros2019review}, cooperative payload delivery \cite{rastgoftar2018cooperative, rossomando2020aerial}, and hazardous environment sensing \cite{argrow2005uav}, several multi-agent coordination techniques have been researched and presented. 

A group of agents, acting as particles of a single virtual rigid body, use the centralized technique of Virtual Structure (VS) \cite{lewis1997high, ren2002virtual}. VS is capable of maintaining the rigid geometric relationship between the agents and evolving as a rigid body in a given direction and orientation. Consensus \cite{ren2005consensus, ding2013network} is among the most exhaustively researched cooperative control approaches. In this approach, a team of agents reaches an agreement or consensus regarding some quantities of interest only by communicating with their neighbors. It is a decentralized coordination and control approach and is broadly divided into two categories: leaderless consensus (i.e., consensus without a leader) \cite{rao2013sliding, ren2009distributed} and leader-follower consensus (i.e., consensus with a leader) \cite{song2010second, xu2016robust}.

Another decentralized leader-follower method is called Containment Control \cite{ferrari2006laplacian, li2013distributed, wang2013distributed, wen2015containment}, where the collective motion of all agents is achieved with multiple leaders. The follower agents obtain
the desired positions through local communication with in-neighbor agents, and all agents are contained within a particular area defined by geometric constraints. A recent multi-agent coordination approach known as Homogeneous (or Affine) Transformation, is based on the principles of continuum mechanics, where the agents in the system are treated as particles of a deformable body undergoing a homogeneous transformation \cite{rastgoftar2014evolution, rastgoftar2016continuum, emadi2021physics, emadi2022collision, emadi2022physics, romano2019experimental, UPPALURU2022107960}. This technique ensures that all agents in the system remain inside a bounding envelope and allows for translation, rotation, and shearing of the bounding envelope while ensuring collision avoidance. Homogeneous transformation advances containment control by ensuring inter-agent collision avoidance. To achieve the desired homogeneous transformation in $n$-D ($n=1, 2, 3$), $n+1$ leaders in $\mathbb{R}^{n}$ communicate with the follower agents via local communication.

\begin{figure}[h]
    \centering
    \subfigure[]{\includegraphics[width=0.99\linewidth]{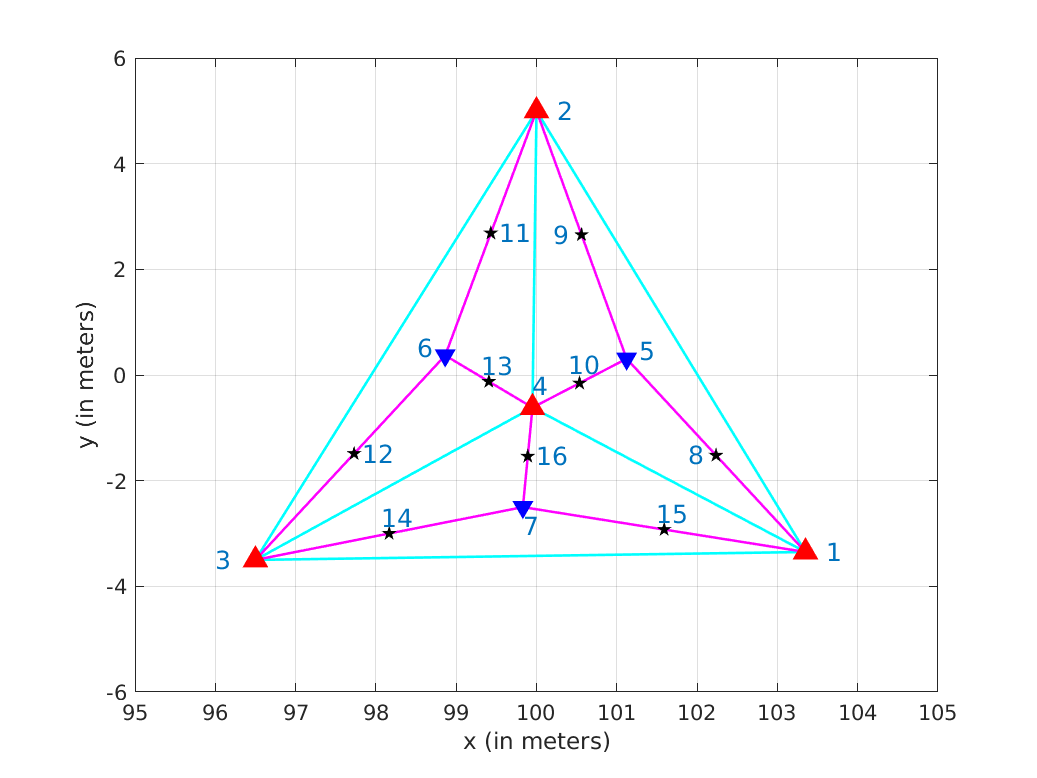}}
    \subfigure[]{\includegraphics[width=0.99\linewidth]{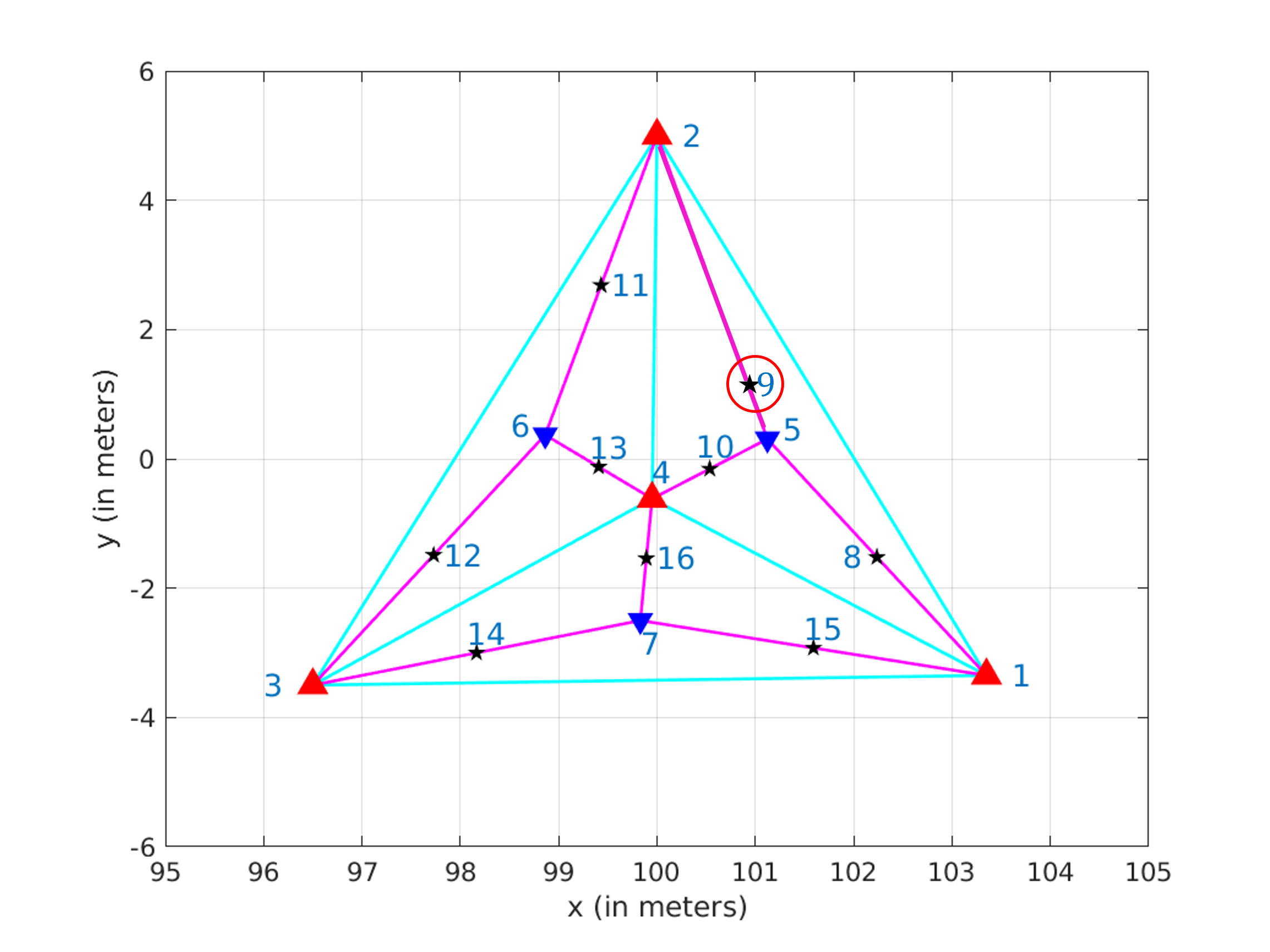}}
    \vspace{.15cm}
    \caption{(a) An example of the reference configuration of a $N = 16$ quadcopter team. As defined in Section \ref{sec:preliminaries}, we have $\mathcal{B}=\left\{1, 2, 3\right\}$, $\mathcal{C}=\left\{4\right\}$ and $\mathcal{I}=\left\{5,\cdots, 16\right\}$. The set $\mathcal{L}_1 = \left\{1, 2, 3, 4\right\}$ is given by red triangles, $\mathcal{L}_2 = \mathcal{L}_1 \cup \left\{5, 6, 7\right\}$ is represented by blue triangles, $\mathcal{L}_3 = \left\{8,\cdots,16\right\}$ is denoted by the black stars. Finally, we have $\mathcal{V} = \left\{1,\cdots,16\right\}$ as the set that identifies all quadcopters in the team. (b) Local deformation of agent $9$ by adjusting weights $\beta_{9,5,3}$ and $\beta_{9,2,3}$.}
    \label{fig:refconfig}
\end{figure}

\subsection{Contributions}
Although homogeneous transformation coordination can allow for aggressive and safe changes to the inter-agent distances, it has ``deformation uniformity'' problems. This is due to the fact that at any moment $t$, the deformation of the complete agent team configuration is given by a single Jacobian matrix that can be specified based on unique rotation and shear deformation angles, as well as axial deformations \cite{rastgoftar2021safe}. Therefore, the rotation, axial, and shear deformations must be consistent over the whole MAS arrangement. To overcome this deformation uniformity issue, the existing homogeneous transformation coordination, specified based on a single Jacobian matrix, is advanced in this paper to Deep Continuum Deformation Coordination (DCDC), which allows us to plan safe ``\textit{local deformation}" of an agent team without having to change the inter-agent distances between all agents ({\color{black}See Fig. \ref{fig:refconfig}(a),(b)}). 

DCDC defines multi-agent coordination as a multi-layer continuum deformation constructed using a neural network in which (i) each neuron represents a different agent and (ii) the input layer receives time-invariant reference positions of the boundary agents. In our proposed NN, which has $p$ hidden layers, the desired deformation of the agents of layer $k+1$ is based on the agents' desired positions belonging to hidden layer $k$ ($k=1,\cdots,p-1$). Unlike existing neural network learning methods that obtain the weights and biases based on the training data, our proposed NN optimization assigns weights and biases based on a single input and a time-varying output. More specifically, our goal is to plan the deformation of a multi-agent team by obtaining the weights and biases of the hidden layers so that the error between the actual position configuration of the agent team and the desired position configuration of the agent team is minimized while guaranteeing inter-agent collision avoidance between every two agents. We provide guarantee conditions for assuring inter-agent collision avoidance by obtaining the inequality and equality constraints on the weights and biases of the hidden layers.

This paper is organized as follows: The preliminaries (Section \ref{sec:preliminaries}) are first introduced, followed by a detailed description of our proposed approach (Section \ref{sec:methodology}). We present safety guarantee conditions (Section \ref{sec:safetyguaranteeconditions}) before presenting simulation results (Section \ref{sec:simulations}). Section \ref{sec:conclusion} concludes the paper.

\section{Preliminaries}
\label{sec:preliminaries}

Consider a $N$-agent team (See Fig. \ref{fig:refconfig}) moving collectively as particles of a deformable body in $3$-D motion space. We use set $\mathcal{V}=\left\{1,\cdots,N\right\}$ to identify all agents in the team. We express set $\mathcal{V}$ as 
\begin{equation}
    \mathcal{V}=\mathcal{B}\bigcup\mathcal{C}\bigcup\mathcal{I},
\end{equation}
where $\mathcal{B}$ defines the primary leader agents that are located at the boundary of the agent team configuration, singleton $\mathcal{C}$ defines a single agent located inside the convex hull defined by $\mathcal{B}$, and  $\mathcal{I}$ identifies the remaining  agents, located inside the convex hull defined by $\mathcal{B}$. Without loss of generality we index agents such that the sets $\mathcal{B}=\left\{1,\cdots,N_L-1\right\}$, $\mathcal{C}=\left\{N_L\right\}$, and $\mathcal{I}=\left\{N_L+1,\cdots,N\right\}$ identify boundary, core, and interior agents, respectively. Set $\mathcal{V}$ is also expressed as
\begin{equation}
    \mathcal{V}=\bigcup_{k=1}^p\mathcal{L}_k
\end{equation}
with subsets $\mathcal{L}_1$ through $\mathcal{L}_p$, where $\mathcal{L}_1=\mathcal{B}\bigcup \mathcal{C}$.  We note that subset $\mathcal{L}_k$ serves as immediate leaders for $\mathcal{L}_{k+1}$, for $k=1,\cdots,p-1$, which implies that desired positions of the agents belonging to $\mathcal{L}_{k+1}$ are defined based on desired positions of the agents in $\mathcal{L}_k$. $\mathcal{L}_2$ through $\mathcal{L}_{p-1}$ are defined as the interior leaders and the set $\mathcal{L}_p$ defines the  pure followers that do not serve as immediate leaders for any agents belonging to $\mathcal{L}_1$ through $\mathcal{L}_{p-1}$. Furthermore, $\mathcal{L}_1$ through $\mathcal{L}_{p-1}$ satisfy the following condition:
\begin{equation}
    \bigwedge_{k=1}^{p-1}\left(\mathcal{L}_k\subset \mathcal{L}_{k+1}\right).
\end{equation}

More specifically, desired position of agent $i\in \mathcal{L}_{k+1}$ is given by
\begin{equation}\label{leaderboundarydesired}
    \mathbf{p}_i(t)=\sum_{j\in \mathcal{L}_k}\beta_{i,j,k}(t)\mathbf{p}_j(t),\qquad i\in \mathcal{L}_{k+1},~k=1,\cdots,p-1,
\end{equation}
where $\beta_{i,j,k}(t)\in \left[0,1\right]$, and
\begin{equation}
    \sum_{j\in \mathcal{L}_k}\beta_{i,j,k}(t)=1,\qquad i\in \mathcal{L}_{k+1},~k=1,\cdots,p-1.
\end{equation}

Safety conditions are explained in Section \ref{sec:safetyguaranteeconditions}. Note that leaders belonging to $\mathcal{L}_1$ are the \textit{primary leaders} that move independently with the desired trajectories defined by
\begin{equation}\label{eq:leaders}
    \mathbf{p}_i(t)=\alpha_{i}(t)\mathbf{a}_{i}+\mathbf{d}(t),\qquad \forall i\in \mathcal{L}_1,
\end{equation}
where $\mathbf{d}(t)$ specifies the nominal position of the agent team configuration expressed with respect to an inertial coordinate system, $\mathbf{a}_i$ is the constant reference position of primary leader $i\in \mathcal{L}_1$. Note that the reference position of the core leader $i\in \mathcal{C}$ is $\mathbf{0}$, i.e. $\mathbf{a}_{i,0}=\mathbf{0}$.

\section{Methodology}
\label{sec:methodology}

\begin{figure}[t]
    \centering
    \includegraphics[width=0.48\textwidth]{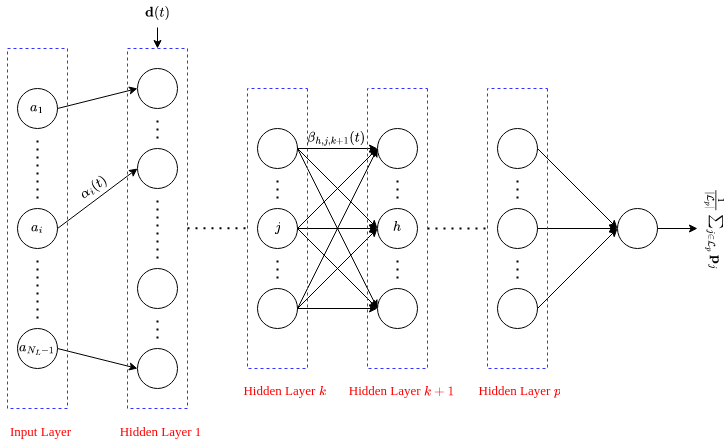}
    \caption{The structure of the proposed neural network used for MQS continuum deformation optimization.}
    \label{fig:NNscheme}
\end{figure}

We investigate deep continuum deformation coordination of a $N$-agent team over a finite time interval $\left[t_0,t_f\right]$ and develop a Neural-Network-based (NN-based) optimization method to obtain the desired  deformation of the agent team (See Fig. \ref{fig:NNscheme}), by assigning $\alpha_i$(t) of the first layer, and $\beta_{i,j,k}$(t) of every layer $k=1,\cdots,p-1$. Figure \ref{fig:NNscheme} illustrates the schematic of the proposed NN with $p$ hidden layers. Note that the artificial neurons in hidden layer $1$ through $p$ represent the agents defined by $\mathcal{L}_1$ through $\mathcal{L}_{p}$, respectively.

\underline{\textit{Input Layer:}} As shown in Fig. \ref{fig:NNscheme}, the input layer generates the reference positions of the boundary agents and core agent, defined by $\mathcal{B} \cup \mathcal{C}$. Note that the reference positions are time-invariant.

\underline{\textit{Hidden Layers:}} The first hidden layer, denoted by $\mathcal{L}_1$, receives the reference positions of the boundary agents. More specifically, neuron $j\in \mathcal{L}_1$ represents agent $j\in \mathcal{B} \cup \mathcal{C}$,  receives reference position $\mathbf{a}_j$, and returns $\mathbf{p}_j(t)$  by using Eq. \eqref{eq:leaders}. Notice that the rigid-body displacement vector $\mathbf{d}(t)$, used in Eq. \eqref{eq:leaders}, has the same bias for all neurons in layer $\mathcal{L}_1$.

Every hidden layer $k\in \left\{2,\cdots,p\right\}$ receives desired positions from previous hidden layer $k-1$ and returns the desired positions of the agents defined by $\mathcal{L}_k$. More specifically,  neuron $i\in \mathcal{L}_{k}$  receives the desired positions of agents of $\mathcal{L}_{k-1}$ and returns $\mathbf{p}_i(t)$ using \eqref{leaderboundarydesired}. Note that the bias of a neuron in hidden layers $2$ through $p$ is $\mathbf{0}$.

\underline{\textit{Output  Layer:}} The output layer consists of a single neuron averaging the desired position of the agent team configuration. Therefore, the weights  associated with the edges connecting neurons of hidden layer $p$ to the output layer are ${1\over \left|\mathcal{L}_p\right|}$ where $\left|\mathcal{L}_p\right|$ denotes the cardinality of set $\mathcal{L}_p$. 

\underline{\textit{Loss function:}} 
In this paper, we use the residual sum of squares as the loss function at time $t$ given by the following equation:
\begin{equation}
    \mathbf{Loss} (t)= \left\|{1\over \left|\mathcal{L}_p\right|}\sum_{i\in \mathcal{L}_p}\mathbf{p}_i(t)-\mathbf{d}\left(t\right)\right\|^2,\qquad t\in \left[t_0,t_f\right].
\end{equation}

\underline{\textit{Training of the Neural Network:}}  To train the network, we determine positive weight parameters $\alpha_i(t)$ and  $\beta_{i,j,k}(t)$ at any time $t\in \left[t_0,t_f\right]$. To ensure safety of the agent team coordination,  $\alpha_i(t)$ and $\beta_{i,j,k}(t)$ are constrained by specific lower and upper bounds that are determined in Section \ref{sec:safetyguaranteeconditions}.

\section{Safety Guarantee Conditions}
\label{sec:safetyguaranteeconditions}

Before proceeding further, we express desired, actual, and reference positions of agent $i\in \mathcal{V}$ as $\mathbf{p}_i(t)=\left[x_{i,d}(t)~y_{i,d}(t)~z_{i,d}(t)\right]^T$, $\mathbf{r}_i(t)=\left[x_i(t)~y_i(t)~z_i(t)\right]^T$, and $\mathbf{a}_{i,0}=\left[x_{i,0}~y_{i,0}~z_{i,0}\right]^T$, respectively. For obtaining the safety guarantee conditions, we make the following assumptions:
\begin{assumption}
Every agent $i\in \mathcal{V}$ can execute a proper trajectory tracking control such that deviation of the actual trajectory $\mathbf{r}_i(t)$ from the desired trajectory $\mathbf{p}_i(t)$ remains bounded and satisfies the following equation
\begin{equation}
    \bigwedge_{i\in \mathcal{V}}\bigwedge_{q\in \left\{x,y,z\right\}}\left|q_{i,d}(t)-q_i(t)\right|\leq \delta,\qquad \forall t,
\end{equation}
where  $\bigwedge_{i\in \mathcal{V}}$ implies ``include all'' and $\delta>0$ is constant. 
\end{assumption}
\begin{assumption}
Every agent $i\in \mathcal{V}$ can be enclosed by a box with side length $2\epsilon$.
\end{assumption}
To assure safety of the agent team continuum deformation, the following two safety requirements must be satisfied:
\begin{safety}
\label{sf1}
For every two different agents $i$ and $j$, the inter-agent collision avoidance is guaranteed, if 
\begin{equation}
\bigwedge_{i=1}^{N-1}\bigwedge_{j=i+1}^N\bigwedge_{q\in \left\{x,y,z\right\}}\left|q_i(t)-q_j(t)\right|>2\epsilon,\qquad \forall t,
\end{equation}
\end{safety}
\begin{safety}\label{sf2}
Every follower $i\in \mathcal{L}_{k+1}$ remains inside the convex hull defined by $\mathcal{L}_k$. We say agents of layer $\mathcal{L}_{k+1}$ are contained by the convex hull defined by $\mathcal{L}_k$, if 
\begin{subequations}
\begin{equation}
    \bigwedge_{k=1}^{p-1}\bigwedge_{i\in \mathcal{L}_{k+1}}\bigwedge_{j\in \mathcal{L}_k}\left(\beta_{i,j,k}(t)\geq 0\right),\qquad \forall t.
\end{equation}
\begin{equation}
    \bigwedge_{k=1}^{p-1}\bigwedge_{i\in \mathcal{L}_{k+1}}\left(\sum_{j\in \mathcal{L}_k}\beta_{i,j,k}(t)=1\right),\qquad \forall t.
\end{equation}
\end{subequations}
\end{safety}
The Safety Condition \ref{sf1} is satisfied, if \cite{rastgoftar2021safe}
\begin{equation}\label{safetysifficient1}
\bigwedge_{i=1}^{N-1}\bigwedge_{j=i+1}^N\bigwedge_{q\in \left\{x,y,z\right\}}\left|q_{i,d}(t)-q_{j,d}(t)\right|>2\left(\delta+\epsilon\right),\qquad \forall t.
\end{equation}
Note that Eq. \eqref{safetysifficient1} provides a sufficient safety condition for inter-agent collision avoidance.

To ensure safety conditions \ref{sf1} and \ref{sf2} are satisfied, we constrain $\alpha_{i,k}$ and  $\beta_{i,j,k}$ by
\begin{equation}
    \bigwedge_{i\in \mathcal{L}_1}\left(\alpha_{min}\leq \alpha_i(t)\right),\qquad \forall t
\end{equation}
\begin{equation}
    \bigwedge_{k=1}^{p-1}\bigwedge_{j\in \mathcal{L}_k}\bigwedge_{i\in \mathcal{L}_{k+1}}\left(\beta_{k,min}\leq \beta_{i,j,k}\leq \beta_{k,max}\right).
\end{equation}
We first assign $\beta_{k,min}$ and $\beta_{k,max}$ for every layer $k=1,\cdots,p-1$. Then, for given $\beta_{k,min}$ and $\beta_{k,max}$ at every layer $k=1,\cdots,p-1$, we obtain $\alpha_{min}$ by solving the following optimization problem:
\begin{equation}
    \alpha_{min}=\min\limits_{\alpha\in \left(0,1\right)}\alpha
\end{equation}
subject to
    \begin{equation}\label{leadersdesired}
        \bigwedge_{i\in \mathcal{L}_1}\bigwedge_{q\in \left\{x,y,z\right\}}\left(q_{i,d}=\alpha q_{i,0} \right),
    \end{equation}
       \begin{equation}\label{followersdesired}
         \bigwedge_{k=1}^{p-1}\bigwedge_{i\in \mathcal{L}_{k+1}}\bigwedge_{q\in \left\{x,y,z\right\}}\left(q_{i,d}=\sum_{j\in \mathcal{L}_k}\beta_{i,j,k}q_j \right),
    \end{equation}
          \begin{equation}\label{mainsafety}
          \resizebox{0.99\hsize}{!}{%
$\begin{split}
                       &\bigwedge_{k=1}^{p-1}\bigwedge_{\substack{i,h\in \mathcal{L}_{k+1} \\ i\neq h}}\bigwedge_{q\in \left\{x,y,z\right\}}\left(\sum_{j\in \mathcal{L}_k}\left|\left(\beta_{i,j,k}-\beta_{h,j,k}\right)q_{j,d}\right|>2\left(\delta+\epsilon\right)\right),\\
                       &\forall \beta_{i,j,k},\beta_{h,j,k}\in \left[\beta_{k,min},\beta_{k,max}\right].
                       \end{split}
                       $
                       }
        \end{equation}
We use Algorithm \ref{euclid33} to solve the above optimization problem.

\begin{algorithm}
  \caption{Obtaining safety guarantee condition for Deep Continuum Deformation Coordination and Optimization}\label{euclid33}
  \begin{algorithmic}[1]
          \State \textit{Get:} $\beta_{k,min}$ and $\beta_{k,max}$ for $k=1,\cdots,p-1$, $\Delta \alpha$, and reference position of every agent $i\in \mathcal{V}$
         \State \textit{Obtain:} $\alpha_{min}$ 
      \State  \textit{Set: $\alpha=1$} 
      \While{Eq. \eqref{mainsafety} is satisfied}
      \State $\alpha\leftarrow \alpha-\Delta \alpha$
      \State Update $\mathbf{p}_i(t)$ of agent $i\in \mathcal{L}_1$ using Eq. \eqref{leadersdesired}
        \For{\texttt{< $ k\in \left\{1,\cdots,p-1\right\}$}}
            \For{\texttt{< $ j\in \mathcal{L}_k$}}
                \For{\texttt{< $ q\in \left\{x,y,z\right\}$}}
                     \For{\texttt{< $ \beta_{i,j,k}\in \left[\beta_{k,min},\beta_{k,max}\right]$}}
                         \State Update $\mathbf{p}_i(t)$ using Eq. \eqref{followersdesired}.
                         \State Check constraint \eqref{mainsafety}.
                    \EndFor
                \EndFor
            \EndFor
        \EndFor    
   \EndWhile
   \State $\alpha_{min}\leftarrow \alpha$
  \end{algorithmic}
\end{algorithm}




\begin{table}[ht]
\begin{center}
\begin{tabular}{ |p{2cm}|p{2cm}|p{2cm}| } 
\hline
\hfil Parameter & \hfil Value &  \hfil Units \\
\hline
\hfil $m$ & \hfil $0.5$ & \hfil $\SI{}{kg}$ \\ 
\hfil $g$ & \hfil $9.81$ & \hfil $\SI{}{m/s^2}$ \\ 
\hfil $l$ & \hfil $0.25$ & \hfil $\SI{}{m}$ \\ 
\hfil $I_r$ & \hfil $3.357\times10^{-5}$ & \hfil $\SI{}{kg m^2}$ \\ 
\hfil $I_x$ & \hfil $0.0196$ & \hfil $\SI{}{kg m^2}$ \\ 
\hfil $I_y$ & \hfil $0.0196$ & \hfil $\SI{}{kg m^2}$ \\
\hfil $I_z$ & \hfil $0.0264$ & \hfil $\SI{}{kg m^2}$ \\ 
\hfil $b$ & \hfil $3\times10^{-5}$ & \hfil $\SI{}{N s^2/rad^2}$ \\
\hfil $k$ & \hfil $1.1\times10^{-6}$ & \hfil $\SI{}{N s^2/rad^2}$\\
\hline
\end{tabular}
\end{center}
\caption{Quadcopter Specifications}
\label{table:quadcopterspecifications}
\end{table}
\section{Simulation Results}
\label{sec:simulations}
Here, we present simulation results obtained using PyTorch\footnote{\url{https://pytorch.org/}} on a desktop running Ubuntu $20.04$ LTS with an Intel i$7$ $11$th generation CPU, an NVIDIA GPU, and $16$ GB of RAM. We consider the evolution of a multi-quadcopter system (MQS), consisting of $N=16$ quadcopters with initial formation at time $t=0$, as shown in Fig. \ref{fig:refconfig}a. The quadcopters in the MQS are similar and are modeled using the dynamics established in \cite{rastgoftar2022real}. The quadcopter parameters originally presented in \cite{gopalakrishnan2017quadcopter} are listed in Table \ref{table:quadcopterspecifications}. 

\begin{figure}
\centering
\includegraphics[width=2.5in]{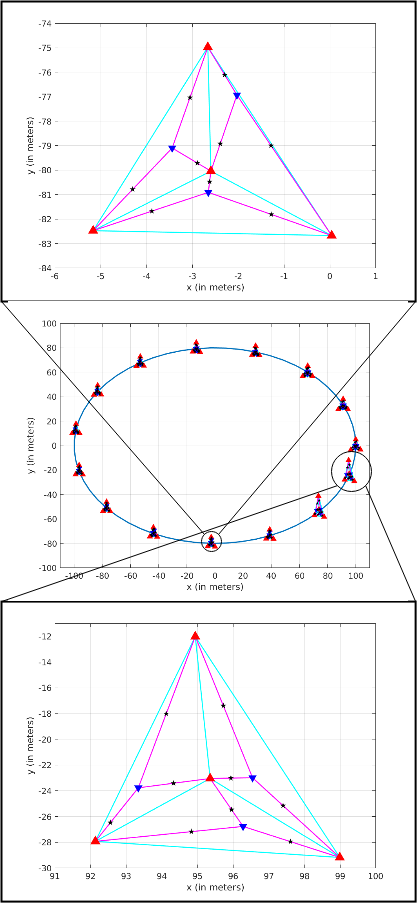}
\caption{Desired configuration of the MQS at various times.}
\label{fig:config_times}
\end{figure}

We consider the desired path of the MQS configuration to be an ellipse as shown in Fig. \ref{fig:config_times} with major radius $a = \SI{100}{m}$ and minor radius $b = \SI{80}{m}$. The total travel time to complete the ellipse is denoted by $T = \SI{60}{s}$. The MQS is distributed over the horizontal plane at $z = \SI{10}{m}$, whose value is always constant. 

The quadcopters' in the MQS are identified by defining the set $\mathcal{V} = \mathcal{B} \bigcup \mathcal{C} \bigcup \mathcal{I}$, where $\mathcal{B} = \{1, 2, 3\}$, $\mathcal{C} = \{4\}$, and $\mathcal{I} = \{5,\cdots,16\}$. Alternatively, as stated in Section \ref{sec:preliminaries}, the sets $\mathcal{L}_1 = \{1,\cdots,4\}$, $\mathcal{L}_2 = \{5,\cdots,7\}$ and $\mathcal{L}_3 = \{8,\cdots,16\}$ identify the quadcopters of hidden layers $1$, $2$, and $3$, respectively. We compute the $\alpha_i$ and $\beta_{i,j,k}$ parameters using the approach presented in Section \ref{sec:methodology}. The minimum and maximum values assigned for the communication weights, $\alpha_i(t)$ and ${\beta}_{i,j,k}(t)$ are shown in Table II. Note that the minimum values are assigned such that inter-agent collision avoidance amongst every two quadcopters in the MQS is guaranteed while also satisfying the constraints mentioned in Section \ref{sec:safetyguaranteeconditions}.

\begin{table}[ht]
    \begin{center}
    \caption{Lower and upper bounds for obtaining the NN weights}
    \begin{tabular}{ccccc} 
    \hline
    \hfil $\alpha_{min}$ & \hfil $\beta_{1,min}$ &  \hfil $\beta_{2,min}$ & \hfil $\beta_{1,max}$ &  \hfil $\beta_{2,max}$ \\
    \hline
    \hfil $0.5$ & \hfil $0.2$ &  \hfil $0.35$ & \hfil $0.6$ &  \hfil $0.65$ \\
    \hline
    \end{tabular}
    \end{center}
    \label{alphabeta}
\end{table}

\begin{figure}[h]
    \centering
    \includegraphics[width=3.3 in]{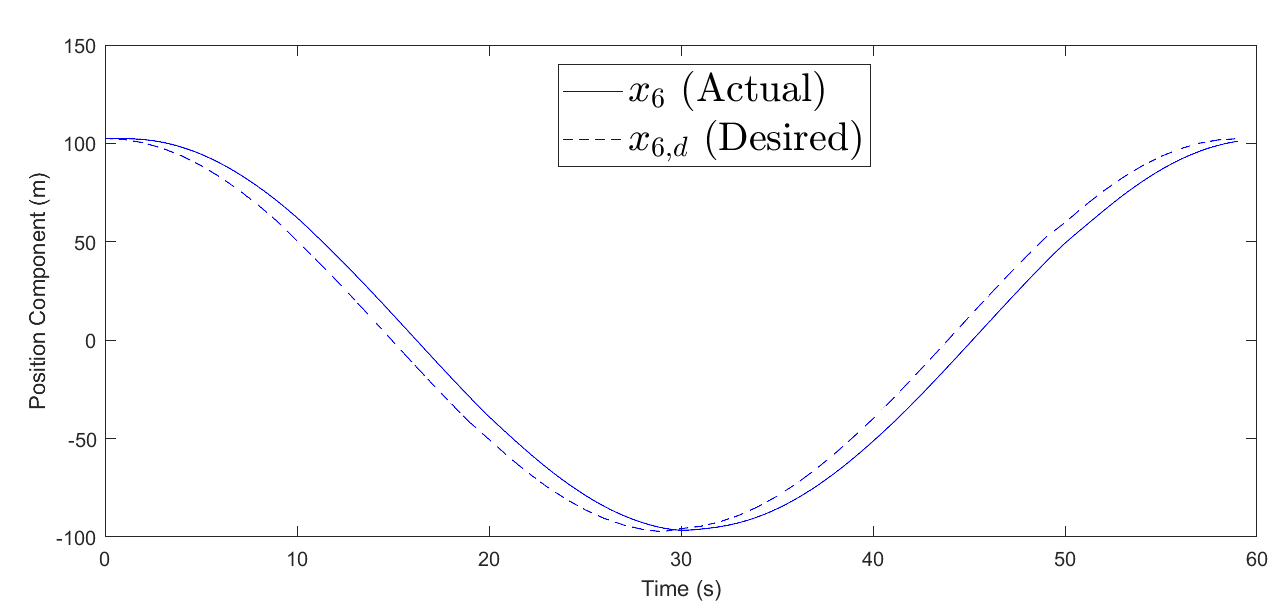}
    \caption{$x$ component of the actual and desired trajectories of quadcopter $6$.}
    \label{x6}
\end{figure}

\begin{figure}[h]
    \centering
    \includegraphics[width=3.3 in]{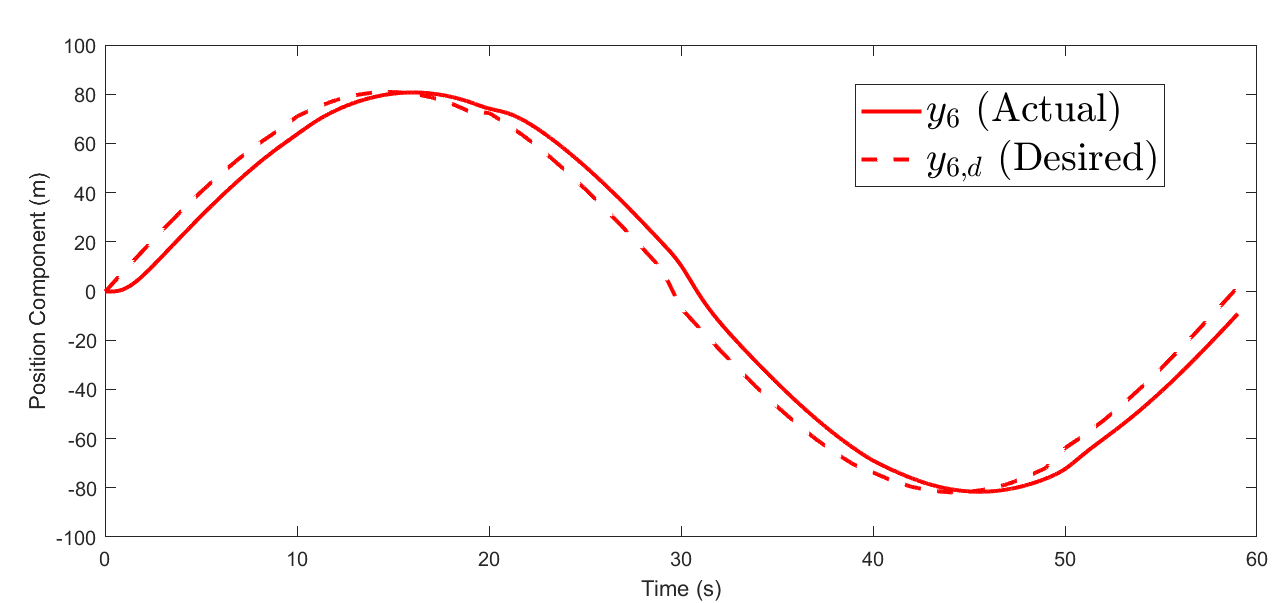}
    \caption{$y$ component of the actual and desired trajectories of quadcopter $6$.}
    \label{y6}
\end{figure}

\begin{figure}[t]
    \centering
    \includegraphics[width=3.3 in]{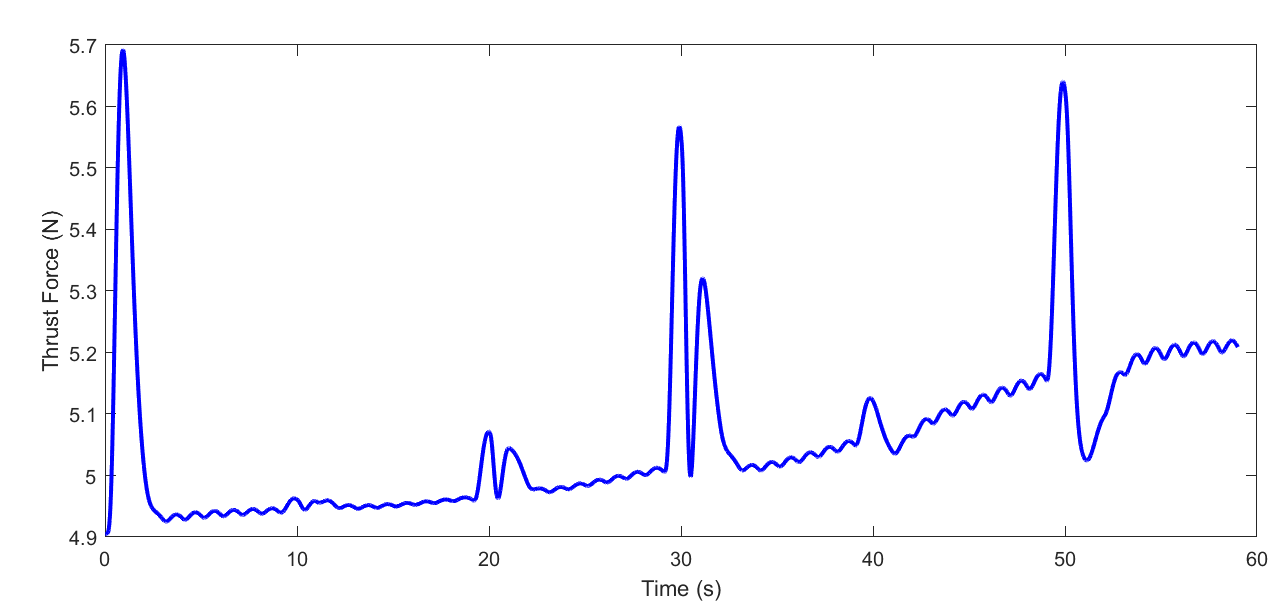}
    \caption{Thrust of quadcopter $6$.}
    \label{p6}
\end{figure}

\begin{figure}[t]
    \centering
    \includegraphics[width=3.3 in]{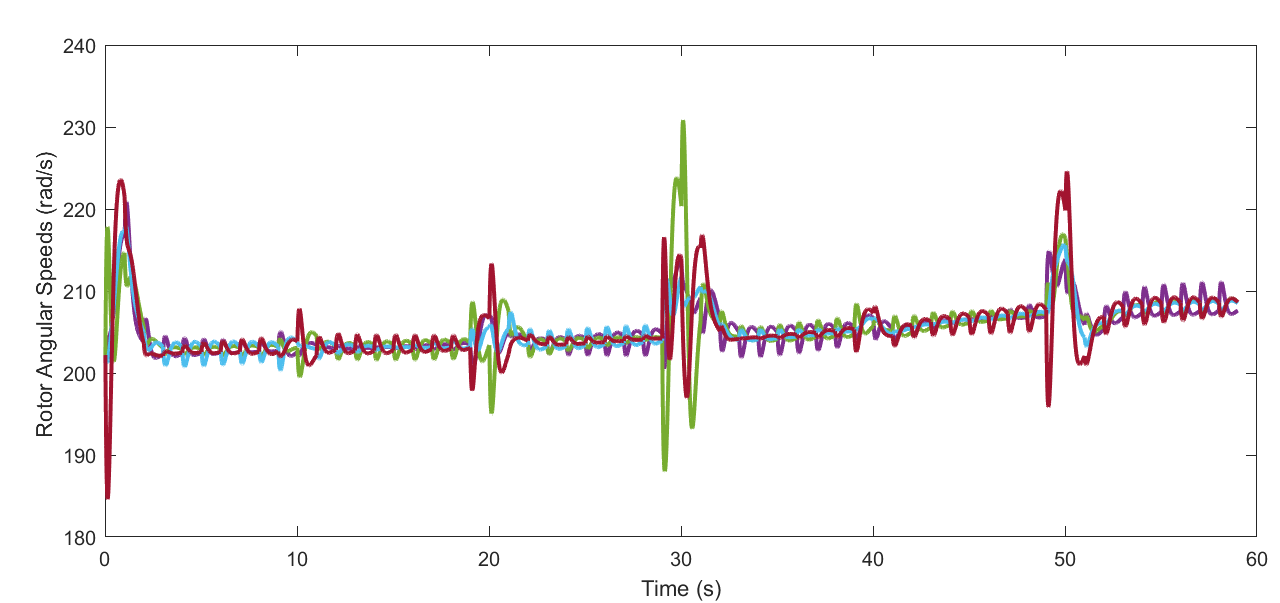}
    \caption{Rotor angular speeds  of quadcopter $6$.}
    \label{s6}
\end{figure}

It is desired that the quadcopter configuration follow the desired elliptic trajectories while satisfying the safety conditions presented in Table II. Therefore, we run the neural network described in Section \ref{sec:methodology} for $6000$ epochs with a learning rate of $0.01$. Stochastic Gradient Descent (SGD) has been used along with the loss function mentioned in Section \ref{sec:methodology}. The weights and biases of the NN are used to calculate the nominal trajectories of each of the quadcopters in the MQS team. Using the input-output feedback linearization control approach \cite{rastgoftar2022real}, Figures \ref{x6}, \ref{y6}, \ref{p6}, and \ref{s6} plot the $x$-position component, the $y$-position component, the thrust force magnitude, and the rotors' angular speeds of the quadcopter $6 \in \mathcal{L}_2$, respectively.  

\section{CONCLUSIONS}
\label{sec:conclusion}

This paper has developed and presented a NN-based technique for safe continuum deformation coordination and optimization of a $N$-quadcopter team tracking a known target trajectory in $3$-D motion space. Assuming that there are no obstacles while tracking the desired trajectory and no failures in the MQS, we presented a novel algorithm in which the input to the NN is the constant reference configuration and optimization is done based on the desired trajectory to track so as to obtain the weights that directly correlate to the matrices in homogeneous transformation. Our work also provided safety guarantees ensuring inter-agent collision avoidance. Future work lies in the direction of making the algorithm robust to obstacles in the environment, conducting further simulations and flight experiments with obstacles, and incorporating fluid flow navigation \cite{emadi2021physics, UPPALURU2022107960, romano2022quadrotor} as an obstacle avoidance algorithm. An alternate direction would be to obtain the deformation of the MQS formation through a known obstacle, such as windows of different shapes and sizes.


\bibliographystyle{IEEEtran}
\bibliography{root}

\end{document}